%% file: sample-sigconf.tex
\documentclass[sigconf]{acmart}
\usepackage{algorithm}
\usepackage{textcomp}
\usepackage{enumitem}
\usepackage{dblfloatfix}
\usepackage{lipsum}
\usepackage{soul}
\usepackage{xcolor}
\usepackage{verbatimbox,caption,float,lipsum}
\newfloat{Code}
\usepackage{algpseudocode}
\usepackage{algorithmic}
\usepackage{wrapfig}
\usepackage{gensymb}
\usepackage{textcomp}
\usepackage{algorithm}
\usepackage{listings}
\usepackage{subcaption}
\usepackage{booktabs}
\usepackage{amsmath}
\usepackage{enumitem}
\usepackage{dblfloatfix}
\usepackage{lipsum}
\usepackage{soul}
\usepackage{xcolor}

\usepackage{algorithmic}
\usepackage{textcomp}
\usepackage{xcolor}
\usepackage{amsfonts}
\usepackage{float}
\usepackage{relsize}
\usepackage{grffile}
\usepackage{algorithm}
\usepackage{algorithmic}
\setcopyright{rightsretained}
\copyrightyear{2023}
\acmYear{2023}
\setcopyright{acmlicensed}\acmConference[RACS '23]{International Conference on Research in Adaptive and Convergent Systems}{August 6--10, 2023}{Gdansk, Poland}
\acmBooktitle{International Conference on Research in Adaptive and Convergent Systems (RACS '23), August 6--10, 2023, Gdansk, Poland}
\acmPrice{15.00}
\acmDOI{10.1145/3599957.3606250}
\acmISBN{979-8-4007-0228-0/23/08}

\begin{document}
\title{Integration of Digital Twin and Federated Learning for Securing Vehicular Internet of Things}

\author{Deepti Gupta}

\affiliation{%
  \institution{Texas A\&M University - Central Texas}
  \streetaddress{1001 Leadership Pl,}
  \city{}
  \state{Texas, USA}
  \country{}
  \postcode{76549}
}
\email{deepti.mrt@gmail.com}

\author{Shafika Showkat Moni}

\affiliation{%
  \institution{Embry-Riddle Aeronautical University}
  \streetaddress{P.O. Box 1212}
  \city{Daytona Beach}
  \state{Florida, USA}
  \country{}
  \postcode{43017-6221}
}
\email{shafika1403@gmail.com}

\author{Ali Saman Tosun}
\affiliation{%
  \institution{University of North Carolina at Pembroke}
  \city{North Carolina, USA}
  \country{}
  }
\email{ali.tosun@uncp.edu}

%\author{Lavanya Elluri}

%\affiliation{%
  %\institution{Texas A\&M University - Central Texas}
 % \streetaddress{1001 Leadership Pl,}
  %\city{}
  %\state{Texas, USA}
 % \country{}
 % \postcode{76549}
%}
%\email{elluri@tamuct.edu}

\begin{abstract}

In the present era of advanced technology, the Internet of Things (IoT) plays a crucial role in enabling smart connected environments. This includes various domains such as smart homes, smart healthcare, smart cities, smart vehicles, and many others. The IoT facilitates the integration and interconnection of devices, enabling them to communicate, share data, and work together to create intelligent and efficient systems. With ubiquitous smart connected devices and systems, a large amount of data associated with them is at a prime risk from malicious entities (e.g., users, devices, applications) in these systems. Innovative technologies, including cloud computing, Machine Learning (ML), and data analytics, support the development of anomaly detection models for the Vehicular Internet of Things (V-IoT), which encompasses collaborative automatic driving and enhanced transportation systems. However, traditional centralized anomaly detection models fail to provide better services for connected vehicles due to issues such as high latency, privacy leakage, performance overhead, and model drift.

Recently, Federated Learning (FL) has gained significant recognition for its ability to address data privacy concerns in the IoT domain. In the context of V-IoT, which involves autonomous vehicles and intelligent transportation systems with connected vehicles communicating with various sensors and devices, FL is used to develop an anomaly detection model. Current technology, the Digital Twin (DT), proves beneficial in addressing uncertain crises and data security issues by creating a virtual replica that simulates various factors, including traffic trajectories, city policies, and vehicle utilization. This enables the system to facilitate efficient and inclusive decision-making. However, the effectiveness of a V-IoT DT system heavily relies on the collection of long-term and high-quality data to make appropriate decisions. Consequently, its advantages may be limited when confronted with urgent crises like the COVID-19 pandemic.

This paper introduces a Hierarchical Federated Learning (HFL) based anomaly detection model for V-IoT, aiming to enhance the accuracy of the model. Our proposed model integrates both DT and HFL approaches to create a comprehensive system for detecting malicious activities using an anomaly detection model. Additionally, real-world V-IoT use case scenarios are presented to demonstrate the application of the proposed model.

\end{abstract}



\maketitle
\input{Introduction}

\input{RelatedWork}
\input{FDL}

\input{SystemModel}

\input{UseCase}
\input{Conclusion}
\bibliographystyle{ACM-Reference-Format}
\bibliography{References}
\end{document}

%% file: Introduction.tex
\section{Introduction}

\begin{figure}
\centering
\includegraphics[width=0.45\textwidth, height=.30\textheight]{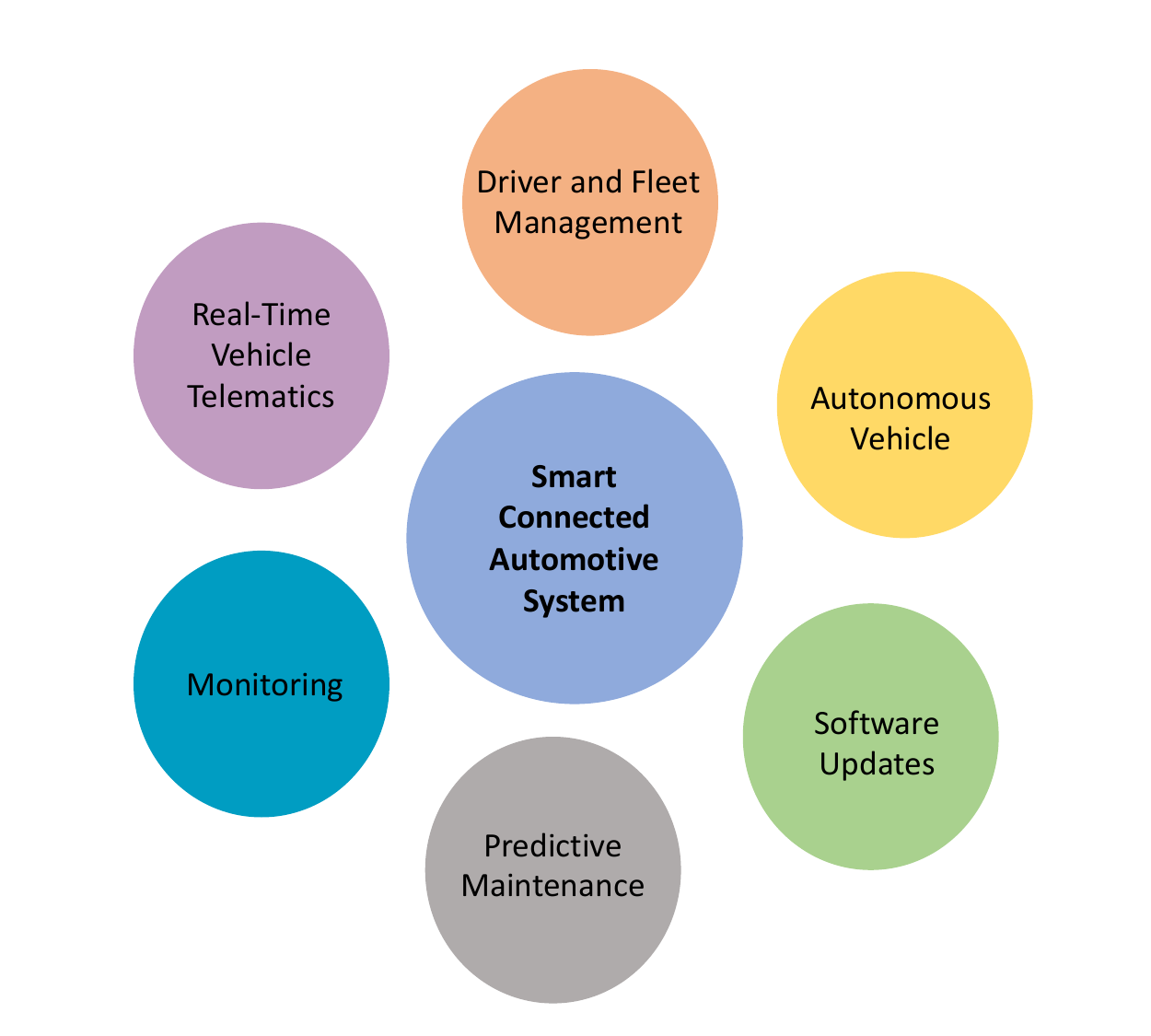}
\centering
\caption{Applications of Smart Connected Automotive System.}
\label{fig:IoV}
\end{figure}

The proliferation of Internet of Things (IoT) technology has become increasingly prevalent in our daily lives, primarily driven by the advancements in low latency and high-speed cellular networks. This technological progress has facilitated seamless connectivity and communication between various smart devices, enabling them to exchange data and interact in real-time. As a result, IoT has found widespread applications in diverse domains such as smart homes, healthcare, transportation, industrial automation, and more, enhancing efficiency, convenience, and automation in our day-to-day activities. As a result of such enormous growth, the number of IoT and connected devices is expected to increase to 60 billion by 2025~\cite{moni2021scalable}. A significant part of this rapid increase is likely linked to the Vehicular Internet of Things (V-IoT). V-IoT comprises connected vehicles, Road Side Units (RSUs), sensors, base stations, edge servers, cloud servers, and other devices capable of data sharing and communication with humans. This information generated through V-IoT will play a vital role in traffic management, traffic safety, infotainment services, smart city, and Intelligent Transportation Systems (ITSs), as shown in Figure\ref{fig:IoV}.

%a DT of a vehicle can collect real-time information e.g. as speed, direction, local traffic information, weather conditions, and information from other DTs of vehicles, and this synchronized information can be utilized to accurately predict future events on roads.

%Federated Learning (FL) is a distributed machine learning approach that prioritizes collaborative learning and privacy without sharing the raw data with a centralized server. FL allows multiple learning agents to collaborate their computing capabilities efficiently and securely to provide better quality of services. Deployment of FL cloud in V-IoT can facilitate the vehicles, RSUs, base station, and other connected devices to to improve the learning efficiency for intelligent environment sensing, intelligent networking, cooperative autonomous driving, and intelligent processing of massive amounts of vehicular data.

Federated Learning (FL) is an innovative approach to Machine Learning (ML) that emphasizes collaborative learning and privacy preservation by avoiding the need to share raw data with a centralized server. In FL, multiple learning agents can efficiently and securely collaborate their computing capabilities to achieve an improved quality of services. The deployment of FL for V-IoT enables vehicles, Roadside Units (RSUs), base stations, and other connected devices to enhance learning efficiency in various aspects such as intelligent environment sensing, intelligent networking, cooperative autonomous driving, and intelligent processing of large volumes of vehicular data. By leveraging FL, V-IoT can benefit from collective intelligence while preserving data privacy and promoting efficient knowledge sharing among the connected entities. 

%add papers based on FL based V-IoT

Recent research studies~\cite{du2020federated, li2023energy, pokhrel2020federated, chai2020hierarchical, lu2020blockchain} have made significant contributions in securing V-IoT environment from anomalies and malicious entities. These studies have proposed and developed various methods and techniques, leveraging the power of FL, to enhance the security and privacy of the V-IoT systems. Additionally, energy-efficient models have been designed, utilizing the FL approach, to optimize energy consumption in V-IoT deployments. The collective findings of these research works contribute to the ongoing efforts in establishing robust security mechanisms and energy efficiency strategies in the V-IoT domain. Despite advances in FL based models for V-IoT system, there are still growing concerns about safety, security and privacy of users. Therefore, this system requires a comprehensive and robust anomaly detection approach to detect anomalous behavior of various entities effectively. 

A Digital Twin (DT) enables connectivity, interaction, and synchronization between the physical entity and its virtual representation in real time. The DT is considered to be one of the most promising technology due to its advanced capabilities, intelligent services, and bridging the gap between the digital model and its physical counterpart. A digital model of a vehicle can be placed at the edge computing node that can serve as an edge middleware in the V-IoT. With the help of this cloud-edge computing paradigm, large-scale data analysis, storage, and modeling are made possible. In addition to that, a huge volume of geographically dispersed information shared by many DTs can be aggregated to derive synthesized information with effectiveness. By creating a virtual replica that mimics real-world conditions, the DT of V-IoT enables simulations and analysis of crucial factors like traffic trajectories, city policies, and vehicle utilization. This virtual representation enhances decision-making processes and assists in developing effective strategies for managing crises while ensuring the security of data within the V-IoT system. 

In this paper, we integrate both DT and FL technologies into the V-IoT framework. Our objective is to harness the advanced computing capabilities offered by DT and leverage the collaborative learning potential of FL to address the security and privacy challenges present in V-IoT systems. By combining these technologies, we aim to enhance the overall performance, efficiency, and privacy preservation in the V-IoT environments. The utilization of DT technology can significantly enhance the efficiency of the anomaly detection model by incorporating data from various sources such as smart sensors, traffic light data, weather statistics, vehicle data, and city policies. The integration of data from multiple vendors enables the provision of more accurate and expedited service delivery for the V-IoT system. In our proposed model, DT facilitates data synchronization and weight aggregation in a synchronous manner, reducing the wait time during FL process. This streamlined approach allows participants to efficiently share their data with the model, ultimately improving the overall performance and effectiveness of the FL-based anomaly detection system. 

In our research, we have implemented a Hierarchical Federated Learning (HFL) approach to develop an anomaly detection model. For example, in \textit{region-1}, where \textit{vendor-1} operates with two smart vehicles, these vehicles collaborate to build their local anomaly detection model. Similarly, in \textit{region-2}, where Vendor-2 operates with five smart vehicles, those vehicles collaborate to construct their own local anomaly detection model. This hierarchical approach allows smart vehicles from multiple regions to collaborate at different levels, enabling the development of robust anomaly detection models for the V-IoT within the same smart city.

By leveraging this HFL approach, our research aims to identify anomalies and enhance the security of the V-IoT systems. Through collaborative learning and data aggregation at different levels, we can improve the accuracy and reliability of the anomaly detection models, ultimately ensuring the integrity and safety of the V-IoT ecosystem. The main contribution of this paper are as follows-
\begin{itemize}

    \item In our research, we have identified a research gap in the development of FL based anomaly detection models specifically tailored for the V-IoT domain.
    
    \item We present the concept of a Hierarchical Federated Learning (HFL) based anomaly detection model.
    
    \item We propose a system model where we integrate both the emerging DT and FL technologies. This approach provides a powerful framework for enhanced collaboration, learning, and decision-making in the V-IoT domain, ultimately leading to improved performance, efficiency, and security.

    \item We also present a use case scenario to demonstrate the feasibility of our proposed model. 

 \end{itemize}
The remainder of this paper is organized as follows. Section~\ref{related} presents the literature review on DT and FL technologies in vehicular internet domains. We also discuss about the V-IoT, DT, and FL in this section. The concept of a Hierarchical Federated Learning (HFL) based anomaly detection model is presented in Section~\ref{FL-VIoT}. Section~\ref{system} presents the proposed system model for identifying anomalies and securing V-IoT. We discuss a use case scenario to demonstrate the practical application of our proposed model in Section~\ref{use case}. Conclusion and future work are discussed in Section~\ref{conclusion}.

%% file: RelatedWork.tex
\section{Related Work and Background}
\label{related}
%This section discusses fundamental concepts and background essential to comprehend the research contributions. It describes the concept of V-IoT, security and privacy issues, DT and FL model.

This section provides an in-depth discussion of fundamental concepts and background information that are essential to understand the research contributions. It covers key topics such as the concept of the V-IoT, security and privacy concerns, as well as FL models and DT.

%\subsection{Related Work}

Recently, there has been a growing interest in various technologies such as cloud computing, edge computing, ML, FL, and DT in both academic and industrial sectors. These technologies are seen as promising solutions for enabling smart cities and ITSs. Lu et al.~\cite{lu2020blockchain} proposed an asynchronous federated framework to implement secure and effective data sharing in the  Internet of Vehicles (IoV). In this approach, each vehicle serves as FL client and shares data with an aggregation server at macro BS (MBS). Vehicles can request a variety of services, including traffic prediction and path selection to the MBS. The MBS develops a shared global model based on accumulated vehicular datasets. Next, the MBS transforms the sharing process into a computing task and resolves the sharing request of vehicles by using an actor-critic reinforcement learning framework.

Chai et al.~\cite{chai2020hierarchical} proposed a hierarchical blockchain-enabled FL scheme for IoV. They presented a feasibility analysis of adapting the hierarchical model to manage large-scale vehicles. In this scheme, each vehicle serves as an FL  client and uses its hardware resources to implement local learning. Road side units (RSUs) are responsible for collecting transactions from vehicles within their communication region in a blockchain framework. Each RSU compute the FL model and append into the blockchain framework to ensure security. The blockchain framework is shared to all RSUs and vehicles in IoV. Shrivastava et al.~\cite{shrivastava2021designing} presented a brief survey on Security in V-Iot using blockchain. In addition, several security models for protecting IoT devices in other domains are discussed in~\cite{gupta2020access, gupta2020learner, gupta2021game, gupta2021future, aslan2021intelligent, ozkan2021comprehensive, gupta2021detecting, gupta2021hierarchical, akbarfam2023dlacb, moni2022secure, pei2022personalized}.

\subsection{Vehicular Internet of Things}
%The vehicular Internet of Things (V-IoT) can be characterized as a platform that facilitates information exchange between a vehicle and its surroundings through vehicle-to-vehicle (V2V), vehicle to infrastructure (V2I), and vehicle to everything (V2X) communication.

V-IoT can be described as a platform that enables the exchange of information between vehicles and their surroundings. This communication is facilitated through various channels, including vehicle-to-vehicle (V2V), vehicle-to-infrastructure (V2I), and vehicle-to-everything (V2X) communication. These interactions allow vehicles to connect with other vehicles, infrastructure elements, and various entities in their environment, creating a networked ecosystem that enhances safety, efficiency, and overall driving experience. Yang et al.~\cite{yang2014overview} put forward an abstract network model for the IoV. Their research focuses on discussing the necessary technologies to establish the IoV framework. They explore various applications that can be built upon existing technologies, highlighting the potential of IoV in different domains. V-IoT enables drivers, pedestrians, and other vehicles to utilize the data produced by vehicular ad hoc networks (VANETs) with the aid of roadside infrastructure ~\cite{moni2020efficient, moni2022crease}. V-IoT integrates the IoT technology with ITSs to improve transportation efficiency and security. It is anticipated that the V-IoT will play a vital role in enabling connected, shared, autonomous, and electric future mobility. This article~\cite{ji2020survey} conducted an extensive literature review focusing on the fundamental aspects of the IoV. They covered essential information related to IoV, including basic VANET technology, different network architectures employed in IoV systems, and typical applications of IoV. 

\begin{figure}
\centering
\includegraphics[width=0.5\textwidth, height=.28\textheight]{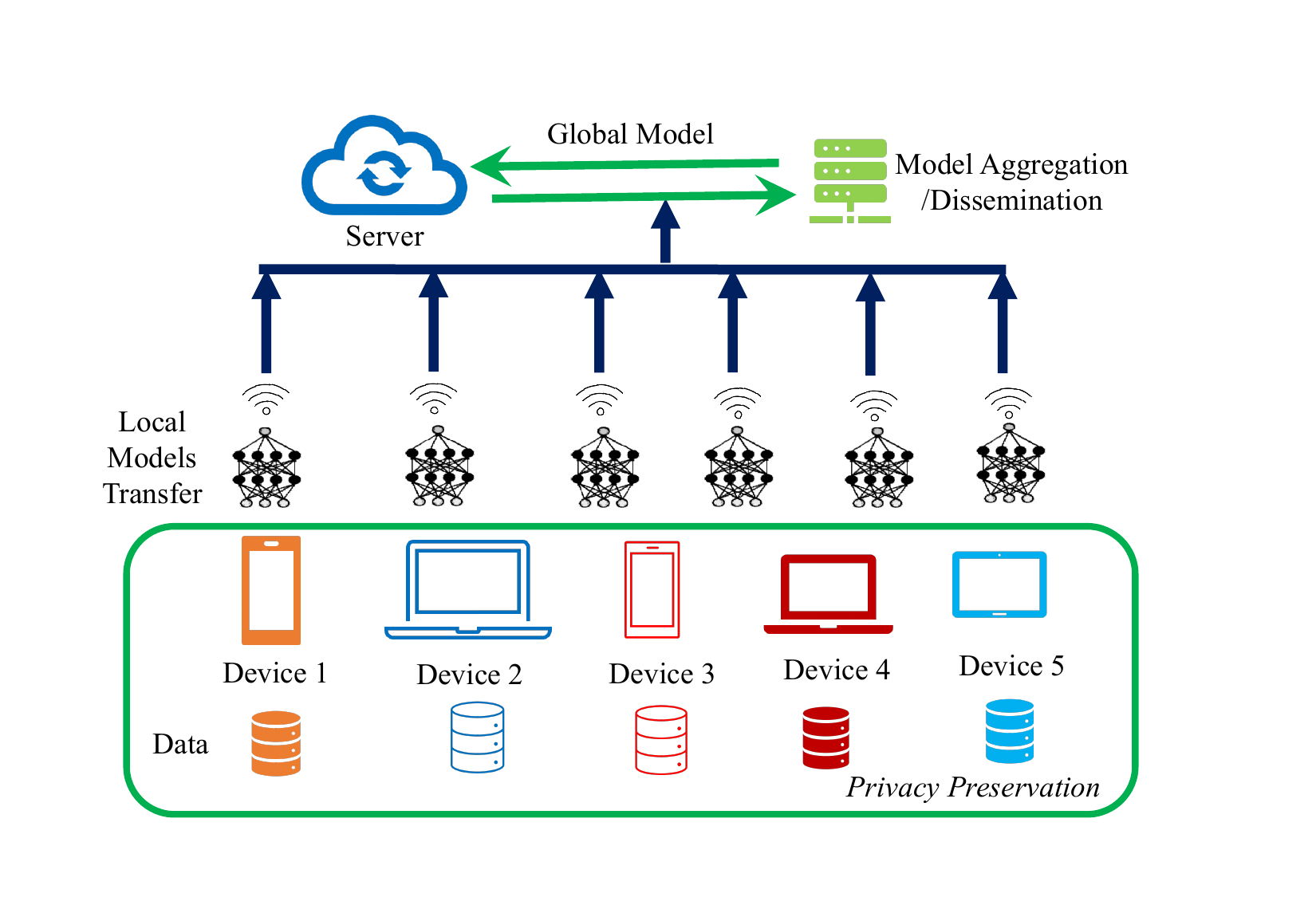}
\centering
\caption{Overview of Federated Learning Model.}
\label{fig:center}
\end{figure}

\subsection{Federated Learning}

FL is an approach that places a strong emphasis on privacy by allowing ML models to be trained locally on individual devices, without the need to share the underlying data with a centralized server. This decentralized training process, depicted in Figure \ref{fig:center}, ensures that sensitive data remains on the devices where it is generated, reducing the risk of privacy breaches. By adopting FL, these privacy risks can be mitigated, as the data remains securely stored on the local devices, and only aggregated model updates are shared with the central server. This way, FL strikes a balance between data privacy and the need for accurate model training, making it a reliable solution for privacy-preserving ML in various applications. Du et al.~\cite{du2020federated} conducted a comprehensive survey of existing studies on FL and its use in wireless IoT. Then, they highlighted the potential benefits of FL in addressing the unique requirements and complexities of vehicular IoT environments. 

Mothukuri et al.~\cite{mothukuri2021federated} introduced a novel approach for anomaly detection in IoT networks using FL. Their proposed method leverages decentralized on-device data to proactively identify intrusions in IoT networks.

%FL is an approach that prioritizes privacy by allowing machine learning models to be trained locally on individual devices without sharing the data with a centralized server, as depicted in the figure \ref{fig:center}. Deep Learning (DL) relies on labeled data to train accurate models capable of performing structured tasks like classification and prediction. However, the process of labeling a dataset comes with an additional privacy cost. Identifying individual data points can have severe consequences for the users who generated that data. The utilization of labeled data can potentially reveal the source, posing risks to security and privacy. In this context, FL has emerged as a reliable solution for preserving privacy in such scenarios.

\subsection{Digital Twin}
DT is referred to as a virtual representation of the real-world entity, devices, machines, process, or other abstraction.  Physical sensors, computer programs, machine learning algorithms, and software models are used to simulate real-time digital models of the physical entity. Tao et al.~\cite{tao2018digital} mentioned that It is considered as one of the most promising enabling technologies for realizing smart manufacturing and Industry 4.0. Wang et al.~\cite{wang2023survey} conducted a comprehensive review of the Internet of Digital Twins (IoDT), focusing on various aspects such as system architecture, enabling technologies, and security/privacy concerns. It facilitates real-time interaction, close monitoring, and reliable communication between the digital model and its physical counterpart. DT is considered to be one of the most promising technology due to its advanced capabilities, intelligent services, and bridging the gap between the digital model and its physical counterpart. For instance, the DT of a vehicle can observe the driving pattern of the driver and communicate, interact, and share this information with other DTs to notify the driver about possible issues or emergencies  on road.

Previous research has introduced several anomaly detection models based on FL in various domains. These models have been deployed either on centralized cloud servers or edge devices. Additionally, the concept of digital twins (DT) has been utilized to identify anomalies in the industrial domain. However, as mentioned earlier, these anomaly detection models often suffer from low accuracy rates due to limited volumes of data available for training. Consequently, a robust anomaly detection model that can provide effective security and privacy solutions for protecting V-IoT systems is still lacking. To bridge this gap, our proposed integrated approach-based anomaly detection model offers a novel perspective for detecting anomalies in the vehicular domain. By combining the strengths of different technologies, such as FL and DT, we aim to enhance the accuracy and effectiveness of anomaly detection in the V-IoT systems. Our approach provides a comprehensive solution that leverages collaborative learning among distributed entities and utilizes the virtual replicas created by digital twins to simulate and analyze various factors contributing to anomalies. We strongly believe that our integrated approach presents a promising solution to address the challenges of anomaly detection in the vehicular domain.
\begin{figure}
\centering
\includegraphics[width=0.5\textwidth, height=.26\textheight]{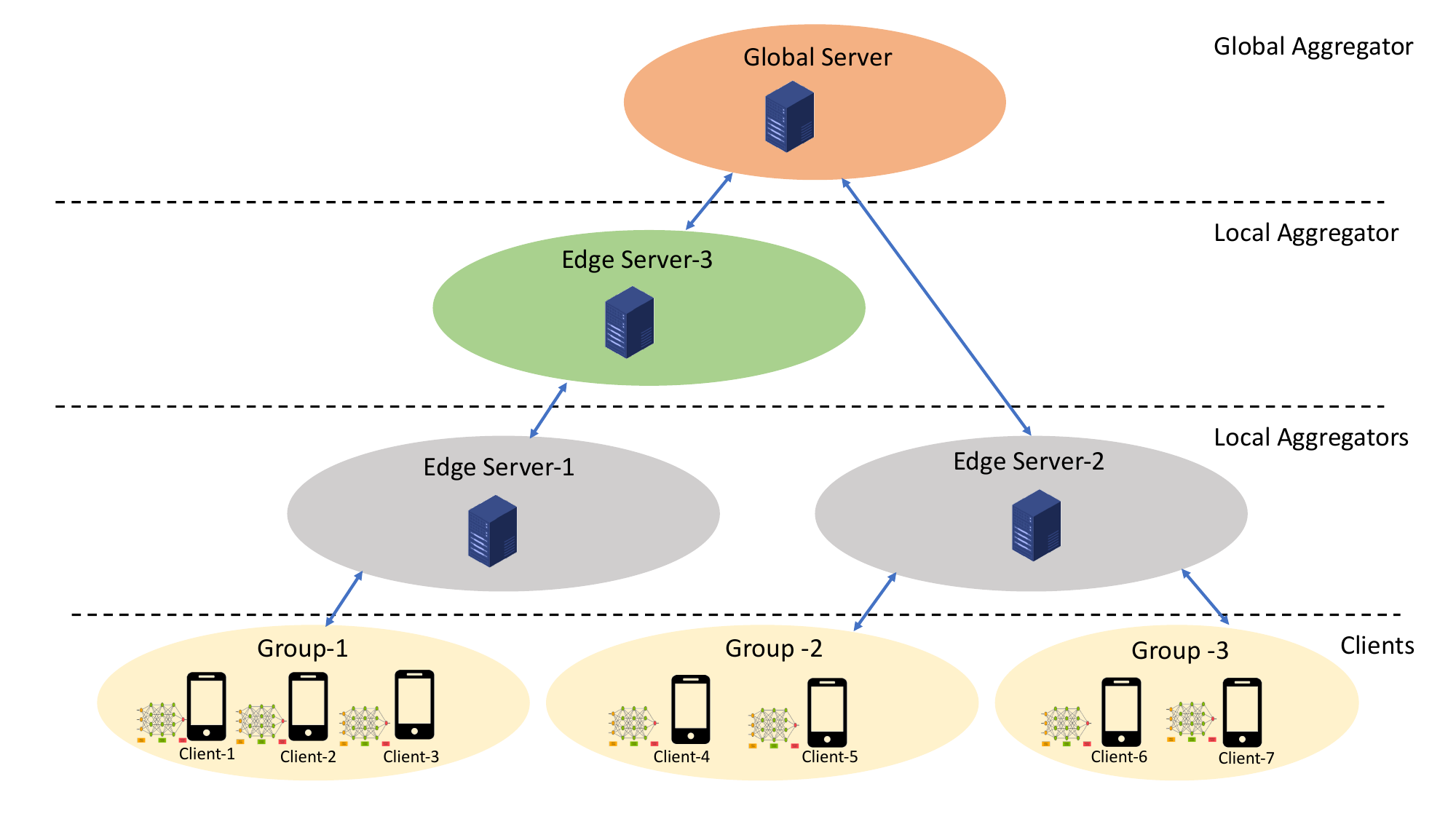}
\centering
\caption{Overview of Hierarchical Federated Learning Model.}
\label{fig:HFL}
\end{figure}

%% file: FDL.tex
\section{Hierarchical Federated Learning based Anomaly Detection Model}
\label{FL-VIoT}

%FL can be widely adopted in the V-IoT domain to train various ML models, including prediction analysis and anomaly detection, by collecting data from vehicles in a privacy-preserving environment. This approach offers potential advantages such as low latency, high efficiency, data privacy, and improved security mechanisms. For the RPM use case, a FL based anomaly detection model \cite{gupta2021hierarchical} is proposed, leveraging edge computing to execute these models locally without sharing patients' data. Subsequently, this research is enhanced for multi-user scenarios by developing a Hierarchical Federated Learning Model (HFL). This model enables the aggregation of gradients at multiple levels to accommodate multiple participants by utilizing the edge computing and DT technologies.. Figure \ref{fig:HFL} presents an overview of this approach. To train the generated V-IoT sensor data, we use \textsc{FedTimeDis} LSTM \cite{gupta2021hierarchical} approach for the connected automotive environment. 

FL can be widely adopted in the V-IoT domain to train various ML models, such as prediction analysis and anomaly detection, by collecting data from vehicles in a privacy-preserving environment. This approach offers several advantages, including low latency, high efficiency, data privacy, and improved security mechanisms. For RPM use case, a FL-based anomaly detection model~\cite{gupta2021hierarchical} is proposed. This model leverages edge computing to execute the anomaly detection models locally on the edge devices without sharing patients' data with a centralized server. To further enhance the capabilities of the FL model for multi-user scenarios, HFL approach is developed. HFL allows the aggregation of gradients at multiple levels, enabling the participation of multiple entities while leveraging edge computing and DT technologies. Figure~\ref{fig:HFL} provides an overview of this approach. In this research, HFL approach is used to develop anomaly detection model for the V-IoT systems.

In the connected automotive environment, there are various types of anomalies, such as traffic congestion, collision detection, malicious attacks, vehicle breakdown, traffic violations and driver fatigue or distraction. The detection and timely response to these anomalies can contribute to improving safety, efficiency, and overall performance in connected vehicle environments. Detecting and understanding anomalies in the V-IoT can lead to enhanced safety, security, performance optimization, and better management of traffic and resources. It allows for proactive decision-making and timely interventions to ensure a smoother and more efficient functioning of the connected vehicle ecosystem.

To develop a HFL based anomaly detection model for the V-IoT, we define the objectives, performance metrics, and requirements for the anomaly detection model. Then, gather relevant data from vehicles in the V-IoT, which may include sensor data, vehicle telemetry, weather statistics, traffic light data and historical records. Ensure that the data collection process preserves privacy and follows ethical guidelines, which is provided by city policies. After that, clean and preprocess the collected data to remove noise, handle missing values, and normalize the features. This step is crucial for preparing the data for further analysis and training. In next step, design the hierarchical architecture for FL in the V-IoT. Where, we need to determine the levels of aggregation such as vehicle-level, region-level, or vendor-level, based on the collaboration requirements and privacy considerations.

To train the data, we utilize the FedTimeDis LSTM~\cite{gupta2021hierarchical} approach, which is specifically designed for the connected automotive environment. Now, each smart vehicle performs local training using their own data. This training is done in a privacy-preserving manner, where data remains on the local device and only model updates (e.g., gradients) are shared. To perform gradient aggregation at each level of the hierarchy to combine the model updates from different participants. This aggregation process ensures that the collective knowledge of the participating entities is utilized to improve the overall anomaly detection model. After developing the model,  evaluate the performance of the aggregated anomaly detection model using evaluation metrics such as accuracy, precision, recall, or F1-score. Refine the model if necessary by adjusting hyperparameters, incorporating feedback, or retraining with additional data. This developed HFL-based anomaly detection model can be deployed in a real-world V-IoT environment and the performance of the deployed model can be monitored continuously. Incorporate new data, update the model periodically, and iterate on the anomaly detection process to enhance its accuracy, efficiency, and robustness.

%Moni please write about smart vehicle sensors, you can tell about V2V,V2I, V2D, After that you can tell about why anomaly detection model is necessary and what types are anomalies we need to identify? tell some examples of anomalies in V-IoT
%The connected automotive environment consists of the 

\begin{algorithm}[!t]
%\SetAlgoLined
\caption{Anomaly Detection of all data nodes $N$}
\label{algo2}
\begin{algorithmic}[1]
\STATE{Collect vehicular the data $D_i$ from all data nodes $N$}
\STATE{Preparing the data $D_i$ by converting into numerical form}
\STATE{Normalize the data $D_i$}
\STATE{Create the sequences sets $(X^n,y^n)$ of data based on correlations}\\
\STATE{Take these input sequence sets $(X^n,y^n)$, where $n= 1,2,$\dots$,N$ from $N$ data nodes, initial model parameter $w_{z}$, local minibatch size $J$, number of local epochs $H$, learning rate $\alpha$, number of rounds $Q$, $h$ hidden layer.}

\STATE{Split local dataset $D_i$ to mini batches of size $J$ which are included into the set $J_i$ and fed horizontally to four LSTM cells.}

\FOR{each local epoch $j$ from 1 to $H$}
\FOR{batch $(X, y)$ $\in$ $J$}
\STATE${h\textsubscript{t} = LSTM(h\textsubscript{t-1}, x\textsubscript{t}, w\textsuperscript{n})}$
\STATE{$y$\textsuperscript{n} = ${\sigma(W\textsuperscript{FC}h\textsubscript{2nd}+Bias)}$}\\
\STATE{$u$\textsuperscript{n} = $w$\textsuperscript{n} - ${w_{z}^n}$ }\\
\STATE{$w_{z}^n$ $\leftarrow$ $w_{z}^n$  + $\frac{\alpha}{N}$ ${\mathlarger{\sum}}\textsubscript{$n$ $\in$ $D_i$} u\textsuperscript{n}$}
\ENDFOR
\ENDFOR 
\STATE{Update weights $w_{z}^n$ to federated cloudlet server and start training again until minimizing the error to build the anomaly detection model.}
\end{algorithmic}
\end{algorithm}

\begin{figure*}[t]
\centering
\includegraphics[width=1\textwidth, height=.44\textheight]{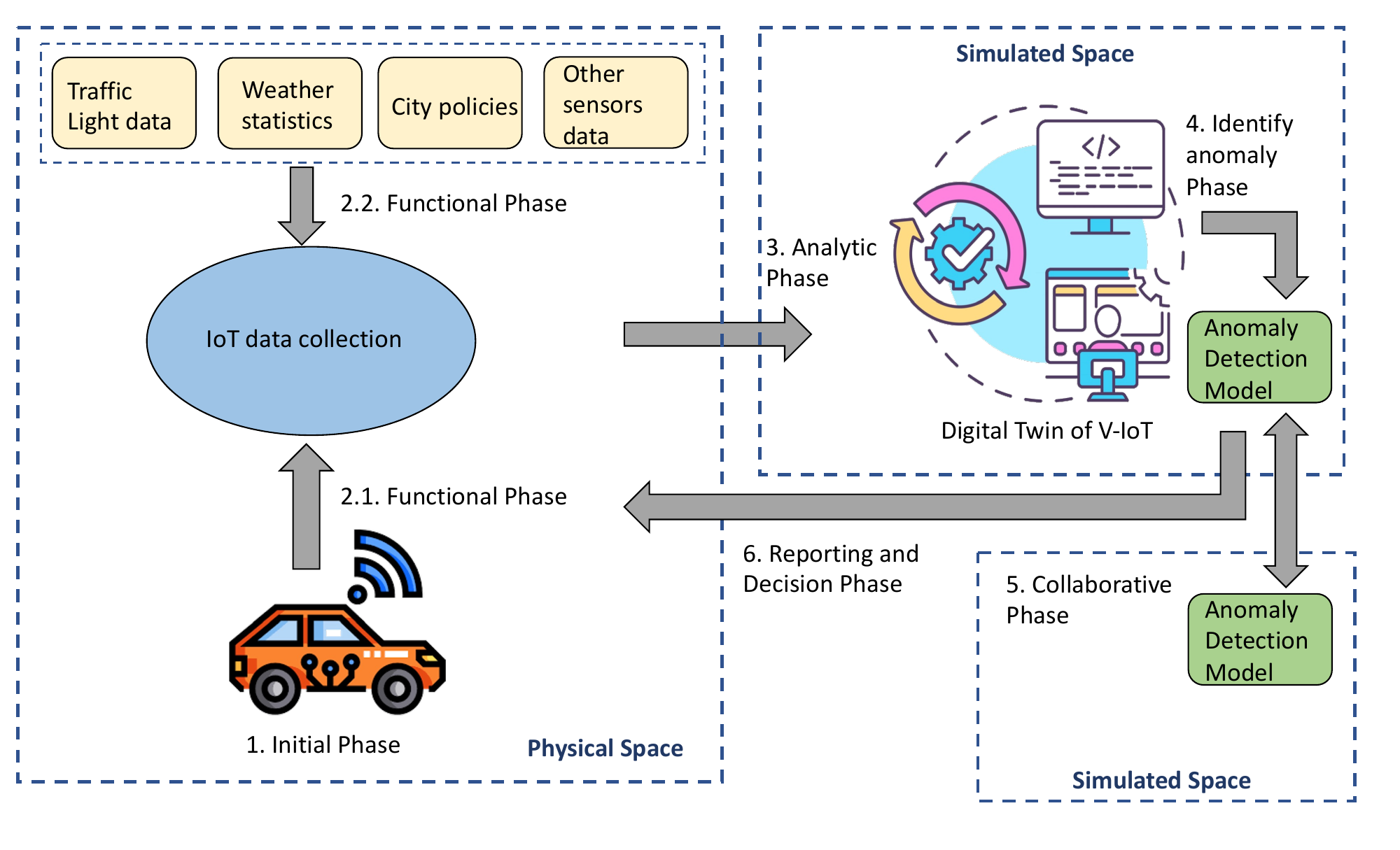}
\centering
\caption{System Model for V-IoT System.}
\label{fig:system}
\end{figure*}

%% file: SystemModel.tex
\section{System Model}
\label{system}

In this section, we introduce our proposed model, which aims to identify anomalies and enhance the security of the V-IoT systems. The model comprises six distinct phases, each involving data exchange and collaboration among different entities. The overall system architecture is illustrated in Figure~\ref{fig:system}, providing a visual representation of the data flow and interaction between the components.

\begin{figure*}[t]
\centering
\includegraphics[width=1\textwidth, height=.44\textheight]{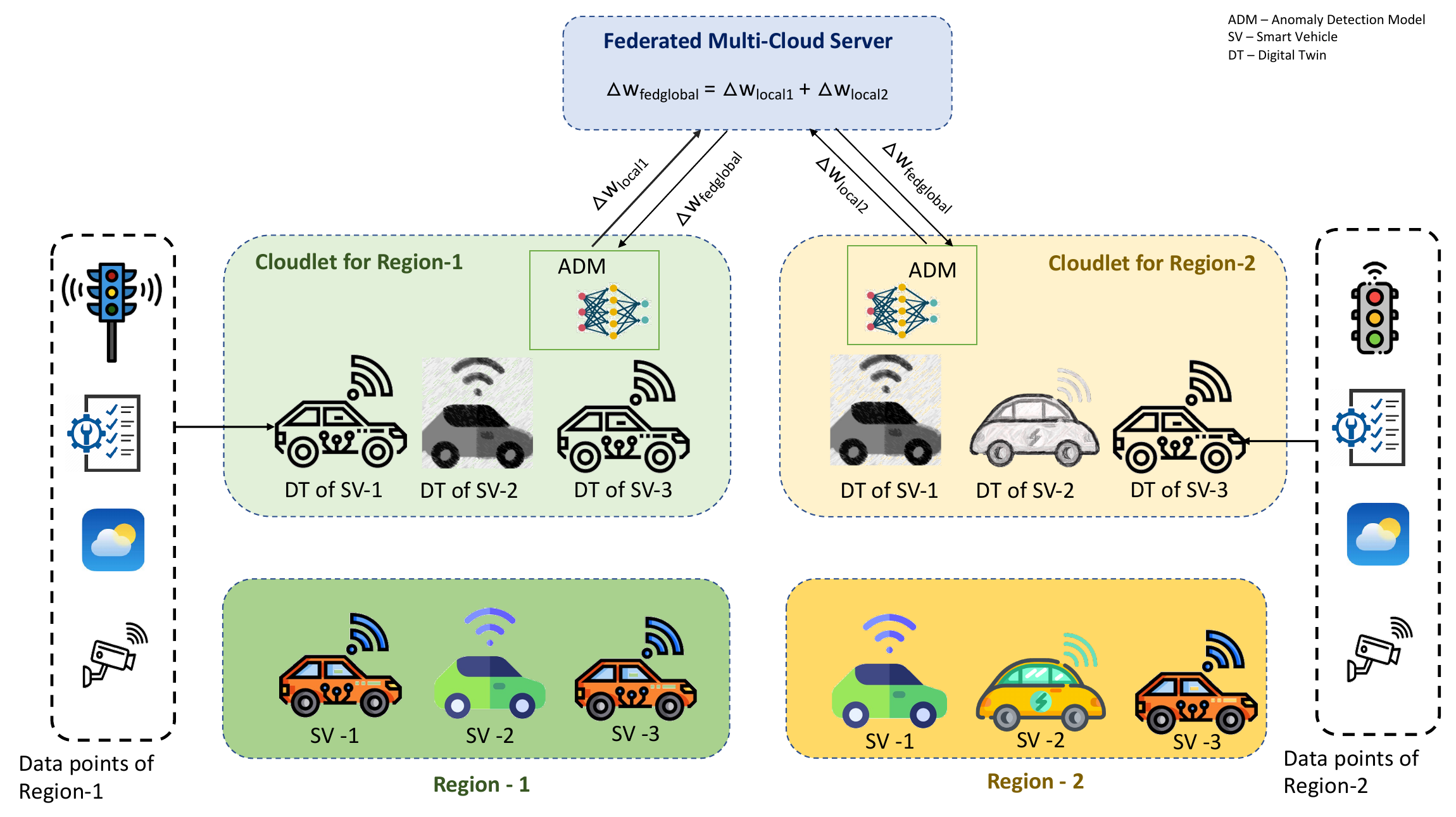}
\centering
\caption{Use Case for Vehicular IoT for Anomaly Detection in Federated Setting.}
\label{fig:anomaly}
\end{figure*}
The six phases of our proposed model are as follows:

\begin{itemize}
    \item \textit{Initial Phase}: During the initial phase of our proposed model, the smart vehicle begins collecting various types of data, including manufacturing data, driver perception data, and external entities data. By collecting these various types of data, the smart vehicle aims to gather comprehensive information about its own performance, the driver's behavior, and the external environment. 

    \item \textit{Functional Phase}: During the functional phase of our model, the entities within the V-IoT system transition into operational mode. This phase is divided into two sub-phases, each serving specific purposes. In the first sub-phase, the focus is on collecting data from the vehicle IoT sensors. These sensors, which are activated and operational, capture various types of information such as vehicle diagnostics, and performance metrics. The collected data is then transmitted to the vendor cloudlet, a cloud-based infrastructure specifically designed to handle V-IoT data. Simultaneously, in the second sub-phase, additional data is collected on the vendor cloudlet. This includes a wide range of data sources such as weather statistics, city regulations and policies, traffic light information, and camera data. These supplementary data sources provide contextual information about the external environment in which the vehicles operate.
    
    By collecting data from both the vehicle IoT sensors and other relevant sources on the vendor cloudlet, a comprehensive and multi-dimensional dataset is created.

    \item \textit{Analytic Phase}: Once the data is collected from the V-IoT system, it is transmitted to the simulated environment for further analysis. This phase involves the transition of data from the physical space to the simulated space, where advanced data analytics techniques are applied. In the simulated environment, a DT is developed for each entity within the V-IoT system. A DT is a virtual representation of a physical entity, in this case, the vehicles and other components of the V-IoT system. The DT is created based on the generated data collected from the previous phases. The data analytics process is then performed on the vehicle DT. Various analytical techniques and algorithms are applied to gain insights and extract valuable information from the data. These analytics help in understanding the behavior, performance, and patterns within the V-IoT system.
    
    By leveraging the DT and conducting data analytics, it becomes possible to identify and understand anomalies within the V-IoT system. Anomalies can include unusual behavior, deviations from normal patterns, or any abnormal activities that may indicate potential security or operational issues.

    \item \textit{Identifying Anomaly Phase}: In this phase, the simulated data $D_i$ from all the data nodes that has been processed and prepared in the previous phase is passed through a pipeline to feed the anomaly detection model. The data is carefully curated and transformed to be compatible with the model's input requirements. The anomaly detection model is developed using suitable machine learning algorithms and follows the Algorithm~\ref{algo2}. These algorithms are trained on the prepared data to learn the patterns and characteristics of normal behavior within the V-IoT system. The model aims to distinguish between normal and anomalous patterns based on the input data. During the training process, the model undergoes iterations to optimize its performance and enhance its ability to accurately detect anomalies. This involves adjusting the model's parameters, fine-tuning the algorithms, and validating the model's performance using appropriate evaluation metrics. Once the training is completed, the anomaly detection model is ready to be deployed and utilized. It collaborates with other models that are part of the subsequent phases, working together to enhance the accuracy and effectiveness of anomaly detection in the V-IoT system.

    %In this phase, the simulated data is passed through the pipeline to feed the anomaly detection model. The data is trained using suitable machine learning algorithms to develop an anomaly detection model. Once the training is completed, the model collaborates with other models that are part of the next phase

    \item \textit{Collaborative Phase}: Indeed, in this phase, the collaboration of multiple anomaly detection models take place to improve the accuracy rate of anomaly detection in the V-IoT system. By combining the weights of multiple ADM models, the overall effectiveness of anomaly detection can be significantly enhanced. Each model may have its own unique approach, algorithm, or specialization in detecting specific types of anomalies. By leveraging the strengths and capabilities of different models, a more comprehensive and robust anomaly detection system can be established. The collaboration among anomaly detection models involves exchanging information, sharing insights, and aggregating their detection results. This collaborative process allows for a holistic analysis of the system's behavior and the identification of anomalies from multiple perspectives.

    %As mentioned, this phase involves the collaboration of multiple Anomaly Detection Models (ADM) to improve the accuracy rate. The ADM models work together to enhance the overall effectiveness of anomaly detection in the system.

    \item \textit{Reporting and Decision Phase}: After an anomalous scenario is detected by the anomaly detection model, it is crucial to report the anomaly to the relevant stakeholders, including the user, vendor, and device. This phase plays a vital role in facilitating informed decision-making and taking necessary actions to ensure the safety and security of the automotive connected environment. Reporting the anomaly to the user is essential as it enables them to be aware of the detected anomaly and take appropriate measures. This could involve alerting the user through notifications, messages, or visual indicators, providing them with information about the anomaly and any recommended actions they should take. Notifying the vendor is also crucial as it allows them to be aware of the anomaly and take the necessary steps to address the issue. This could involve investigating the root cause of the anomaly, analyzing the data collected, and implementing corrective measures to prevent similar anomalies in the future.
    
    Overall, this phase of reporting anomalies is a critical component of the anomaly detection process in the V-IoT system. It helps to minimize the risks associated with anomalous events, enables proactive decision-making, and contributes to maintaining a safe and reliable automotive ecosystem.

    %After an anomalous scenario is detected using the Anomaly Detection Model (ADM), this phase involves reporting the anomaly to the user, vendor, and device. By notifying all relevant parties, this phase enables informed decision-making to ensure the safety of the automotive connected environment.
\end{itemize}

By employing our proposed system model, the V-IoT environment can detect anomalies in real-time and enable prompt responses to the system. This improves overall security and privacy, enhances the efficiency of the transportation system, and improves the driving experience for individuals.

The following section presents a use case scenario of V-IoT, where our proposed system model is employed to detect anomalies.

%% file: UseCase.tex
\section{Use Case Scenario}
\label{use case}
V-IoT is unfolding in many ways where users receive better services and take advantage of autonomous vehicle. The proposed system model where we present the integration of FL and DT, which can be used to secure V-IoT applications, for example, ITSs, cooperative autonomous driving, connected car services, smart city integration, collision avoidance systems and intelligent traffic control etc. In this section, we present a use-case scenario for securing V-IoT by developing an anomaly detection model. The Figure~\ref{fig:anomaly} shows the use case based on our proposed system model, which is discussed in Section~\ref{system}. 

In a smart city, the adoption of IoT technologies and the deployment of smart vehicles are guided by common city policies and regulations. These policies ensure uniformity and standardization across different regions within the smart city. Each region within the smart city may have multiple vendors launching their smart vehicles, contributing to the overall intelligent transportation ecosystem. The presence of multiple vendors in different regions allows for a diverse range of smart vehicles with varying features, technologies, and capabilities. These vehicles may be equipped with advanced sensors, communication systems, and intelligent algorithms to enhance their functionality and contribute to the overall smart city objectives.

In Figure~\ref{fig:anomaly}, the depicted scenario showcases the presence of two different vendors, namely \textit{vendor-1} and \textit{vendor-2}, operating within \textit{region-1} of the V-IoT system. \textit{Vendor-1} has two smart vehicles, SV-1 and SV-3, while \textit{vendor-2} has one smart vehicle, SV-2. Additionally, \textit{region-1} consists of various data points, including sensor data, city policies and regulations, traffic lights, and weather statistics. To enable efficient data processing and anomaly detection in \textit{region-1}, a cloudlet is launched specifically for this region. The cloudlet serves as a localized computing resource that can host and deploy DTs of each smart vehicle within the region. These DTs not only receive data from their respective smart vehicles (SV-1, SV-2, and SV-3) but also incorporate data from other sources within the region, such as sensor data, city policies and regulations, traffic lights, and weather statistics.

In this use case, we deploy DTs on the cloudlet to reduce the gap between physical objects and their digital representations which are generally hosted in the cloud servers. These cloudlets are hosted by the regions and basically a regional cloud that can better cloud services nearer to the user. Cloudlets~\cite{saini2022chapter} are small-scale, mobility-enhanced cloud data centers that sit at the network's edge. The cloudlet's primary goal is to support furious resource and interactive mobile applications by delivering strong computing resources to mobile devices with reduced latency. A wireless local area network with single hop at comparatively higher speed, allows User Equipments (UEs) to connect to the computing resources in the neighboring cloudlet. By leveraging the cloudlet in \textit{region-1}, the DTs can effectively analyze and process the combined data from multiple sources. This integration of data from various entities allows for a holistic understanding of the V-IoT environment within \textit{region-1}. The DTs can leverage this comprehensive data to enhance anomaly detection capabilities, identify patterns, and detect any deviations or anomalies in the behavior of the smart vehicles or the overall V-IoT system. The deployment of DTs within the cloudlet of \textit{region-1} enables localized processing and analysis, reducing latency and enhancing real-time anomaly detection. The collaboration of the DTs with the cloudlet infrastructure facilitates efficient information exchange and enables timely response to any detected anomalies. 

DT of SV-1 within \textit{region-1} plays a crucial role in the development of the anomaly detection model. The data collected by SV-1's DT is utilized to train the initial anomaly detection model specific to \textit{region-1}. This model focuses on detecting anomalies within the local context and behavior of the vehicles and infrastructure within \textit{region-1}. To enhance the accuracy and effectiveness of the anomaly detection model, collaboration is encouraged among multiple models. In this case, the anomaly detection models of \textit{region-1} have the capability to collaborate with each other using the concept of FL. FL allows the models to share their knowledge and insights while maintaining data privacy and security. By aggregating the local models' learnings through weight aggregation techniques, a more robust and accurate anomaly detection model can be obtained.

Moreover, collaboration is not limited to models within the same region. The anomaly detection model of \textit{region-1} can also collaborate with the anomaly detection model of \textit{region-2} by using HFL concept. This collaboration is facilitated by exchanging gradients on a federated multi-cloud server. The gradients represent the model parameters that are shared and utilized to improve the models' performance collectively. By leveraging the collaboration capabilities of FL and the exchange of gradients on the federated multi-cloud server, the anomaly detection models of different regions can benefit from each other's insights and experiences. This cross-region collaboration enhances the overall effectiveness of anomaly detection in the V-IoT system by incorporating knowledge from diverse geographical areas and vehicle behaviors.

In summary, the integration of FL and the exchange of gradients enable collaboration among anomaly detection models at different levels by utilizing DT. This collaboration improves the accuracy and robustness of the models, both within the same region and across different regions, leading to more effective anomaly detection in the V-IoT system.

%% file: Conclusion.tex
\section{Conclusion and Future Work}
\label{conclusion}

In this paper, we have discussed the relevance of FL in V-IoT environments and its potential impact. We started by providing background information on V-IoT, FL, and DT technologies. We emphasized the importance of anomaly detection models in the V-IoT domain and also presented the outline of HFL based anomaly detection model. To address the challenges and opportunities in the V-IoT, we proposed a system model that integrates DT and FL. This model leverages the collaborative learning capabilities of FL and the bridging capabilities of DT between the physical system and its virtual representation. We outlined the key components and phases of our proposed model, emphasizing data exchange, anomaly detection, and security.

To illustrate the practical application of our proposed model, we presented a use case scenario in which the model is employed to detect anomalies in the V-IoT environment. The scenario showcased the data collection, processing, anomaly detection, and collaborative response aspects of our model, highlighting its potential benefits in ensuring safety and efficiency in the V-IoT systems.

Overall, this work aims to contribute to the advancement of FL, DT, and V-IoT research. By introducing our proposed model and presenting a use case scenario, we provide a foundation for further exploration, development, and practical implementation of HFL and DT in the V-IoT environments. We believe that this paper will facilitate the progress of these fields and stimulate further research in the area of HFL-based anomaly detection by utilizing DT in the V-IoT.

%With the increasing interest in FL from various domain, a discourse on the utilization of FL in V-IoT environments becomes significant. In this paper, we have examined the background of V-IoT, FL, and DT. We have highlighted the significance of HFL and presented a system model incorporating DT and FL. Furthermore, we have presented a use case scenario to demonstrate the practical application of our proposed model. We believe that this work has the potential to accelerate the research process for FL, DT and V-IoT. By introducing the proposed model and presenting its use case scenario, we aim to contribute to the advancement of these fields and provide a foundation for further exploration and development.